\documentclass[aps,pra, twocolumn, superscriptaddress]{revtex4-1}

\pdfoutput=1

\usepackage{hyperref,graphicx,amsmath,latexsym,revsymb,amssymb,verbatim,color}

\newenvironment{proof}[1][Proof]{\begin{trivlist}
\item[\hskip \labelsep {\bfseries #1}]}{\end{trivlist}}

\begin{document}

\title{Odd orders in Shor's factoring algorithm}

\author{Thomas Lawson} \affiliation{LTCI -- T\'el\'ecom ParisTech, 23 avenue d'Italie, 75013, Paris, France}

\date{\today}
\begin{abstract}
Shor's factoring algorithm (SFA) finds the prime factors of a number, $N=p_1 p_2$, exponentially faster than the best known classical algorithm. 
Responsible for the speed-up is a subroutine called the \emph{quantum order finding algorithm} (QOFA) which calculates the \emph{order} -- the smallest integer, $r$, satisfying $a^r \mod N =1$, where $a$ is a randomly chosen integer coprime to $N$ (meaning their greatest common divisor is one, $\gcd(a, N) =1$). 
Given $r$, and with probability not less than $1/2$, the factors are given by $p_1 =  \gcd (a^{\frac{r}{2}} - 1, N)$ and $p_2 = \gcd (a^{\frac{r}{2}} + 1, N)$. 
For odd $r$ it is assumed the factors cannot be found (since $a^{\frac{r}{2}}$ is not generally integer) and 
the QOFA is relaunched with a different value of $a$. 
But a recent paper
[E. Martin-Lopez \emph{et al.}: Nat Photon {\bf 6}, 773 (2012)]
noted that the factors \emph{can} sometimes be found from odd orders if the coprime is square. 

This raises the question of improving SFA's success probability by considering odd orders. 
We show that an improvement is possible, though it is small. We present two techniques for retrieving the order from apparently useless runs of the QOFA:  
not discarding odd orders; 
and looking out for new order finding relations in the case of failure. 
In terms of efficiency, using our techniques is equivalent to avoiding square coprimes and disregarding odd orders, which is simpler in practice. 
Even still, our techniques may be useful in the near future, while demonstrations are restricted to factoring small numbers. The most convincing demonstrations of the QOFA are those that return a non-power-of-two order, making odd orders that lead to the factors attractive to experimentalists. 
\end{abstract} 

\maketitle
\section{Introduction}
The most famous application of quantum computers is Shor's factoring algorithm (SFA), which promises to factor a number, $N$, in time $O( (\log N )^3)$, 
much faster than the best known classical routines whose run time increases exponentially in the length of $N$. 
SFA has been extensively studied theoretically, but it has not yet been convincingly demonstrated in the lab; 
the difficulty of controlling quantum systems means just a handful of experiments have been done, to test the basic principles \cite{eml-nphot-6-773,Vandersypen2001,lu-prl-99-250504,Lanyon2007,Mathews2009,lu-nphys-10-719}. 
These experiments are too simple to be of practical use but are, nonetheless, important. 
They have revealed previously unappreciated quirks of the algorithm, one of which -- the role of odd orders -- is the subject of this letter. 

Much of SFA can be done quickly on a classical computer: the Euclidean algorithm lets one pick the coprime, $a$, at random, and calculate the factors given $r$. 
The part which is slow classically -- and speeded-up by quantum mechanics -- is the process at the heart of SFA: calculating $r$. 
The \emph{quantum order finding algorithm} (QOFA) 
uses phenomena such as quantum superposition and entanglement to calculate $r$ efficiently. 
Even though in practice this is the hardest part of SFA to build, it is not the only source of failure. 
Sometimes, despite the QOFA finding $r$ correctly, the classical algorithm does not return $p_1$ and $p_2$. 
(We assume that the QOFA returns $r$ with certainty, although in practice the QOFA will occasionally fail to find $r$, either though experimental error, or because the continued fractions algorithm has not worked.)
Given $r$, failure occurs when the QOFA returns either the trivial factors, $1$ and $N$, or an odd value of $r$, in which case $a^{\frac{r}{2}} \pm 1$ is not generally integer and the QOFA is relaunched with a different value of $a$ \cite{nielsen+chuang2000}.

However, a recent paper \cite{eml-nphot-6-773} (to which the author contributed) noted that the factors \emph{can} sometimes be found from odd values of $r$, if $a$ is square. 
This is interesting for two reasons. 
First, 
it contradicts the almost universally held belief that odd orders are not useful:
every description of SFA specifies that odd orders should be disregarded  (see, for example, reference \cite{nielsen+chuang2000}).

Second, considering odd orders may improve the success probability of factoring. 
Several studies have considered modifying the classical part of the algorithm so as to speed up SFA \cite{Leander2002,Markov2012,Shor1994,Knill1995}. 
The goal is generally to reduce the dependence on quantum processing by replacing it with a classical computation. 
The benefits of this technique are generally underestimated when viewed purely in terms on efficiency. Each quantum circuit, being a physical experiment, must be constructed in the laboratory, a process which is slow, and costly in resources; for some architectures, photonics, for instance, a new circuit must be built for each calculation.  
A common strategy is to reduce the probability of finding useless orders which, though not fatal to the exponential speed-up, contribute significantly to the run-time of SFA. 
So far, studies doing this have concentrated on reducing the occurrence of odd orders, assuming them not to be useful \cite{Leander2002,Markov2012}. 
The most successful of these is reference \cite{Leander2002}, which shows that odd orders can be avoided by picking a coprime that is non-square under modular arithmetic, a property one can efficiently check using the Jacobi symbol. 

Knowing that odd orders \emph{can} in fact be useful, we reverse this logic, asking whether
the success rate of SFA is improved by \emph{considering} odd orders. 
For instance, it is conceivable that finding the factors is easier using square coprimes, which precipitate the useful kind of odd orders. Such as result would boost the efficiency of SFA significantly. 
Alas, as we show, this is not the case. 
Nonetheless, a small improvement is possible. 
By presenting two techniques -- considering odd orders, and checking for new order-finding relations which can be a consequence of failure -- we show that the factors can be found from a square coprime, providing they would have been found from its (non-square) root, for which it is not necessary to consider odd orders. 
In other words, if the goal is to improve the success rate of SFA, simplest is to avoid square coprimes rather than to consider odd orders; 
despite a different starting point, we reach the same conclusion as reference \cite{Leander2002}. 
The fact that randomly picked coprimes are unlikely to be square, especially for large $N$, means the improvement is small (smaller than that proposed in reference \cite{Leander2002}, which benefits from using the stronger property of non-squareness under modular arithmetic). 

Even still, in practice our techniques for using odd orders may be useful, especially in the near future. 
Demonstrations of the QOFA are more convincing when the order is not a power of two, $r\neq 2^p$ for integer $p$, since the output of such an experiment is sensitive to imperfections throughout the circuit \cite{Smolin2013,eml-nphot-6-773}. In contrast, the output of a experiment returning $r=2^p$ matches that of a malfunctioning, non-entangling circuit, making it hard to know if the circuit is working correctly. 
While imperfect technology restricts the size of demonstrations, odd orders that lead to the factors are particularly attractive, accounting for many of the orders of the form $r\neq 2^p$ for small $r$. 
This is precisely why Martin-Lopez \emph{et al.} \cite{eml-nphot-6-773} considered factoring $N=21$ using square coprime $a=4$, giving order $r=3$: it is the simplest calculation which tests the efficacy of the quantum circuit, but which still leads to the factors.

\section{Order finding}
Before investigating the role of odd orders, we review the classical part of SFA, showing where the factors come from. 

By assumption, $r$ is the smallest integer that respects $a^{r}\mod N =1$,
or, equivalently,
\begin{align}
N | \big(a^{\frac{r}{2}} -1\big) \big(a^{\frac{r}{2}} +1\big),
\end{align}
where $|$ means \emph{divides}. 
Assuming that $a^{\frac{r}{2}}$ is integer, the factors can be found provided two conditions are met, 
\begin{align}
N \nmid & a^{\frac{r}{2}} -1 \label{eq: criteria1}\\
N \nmid & a^{\frac{r}{2}} +1, \label{eq: criteria2}
\end{align}
where $ \nmid $ means \emph{does not divide}. 
If so, $a^{\frac{r}{2}} \pm 1$ must each be divisible by one of the factors of $N$ 
and, hence, the factors are $p_1 =  \gcd (a^{\frac{r}{2}} - 1, N)$ and $p_2 = \gcd (a^{\frac{r}{2}} + 1, N)$. 

If one of the conditions \eqref{eq: criteria1} or \eqref{eq: criteria2} is not met, a factor will not be found, except if it is equal to one, two or $N$, in which case $N$ is either even (and is, thus, easy to factor without a quantum computer) or one of the trivial factors has been found. 
For instance, the factor $p_1=\gcd(a^{\frac{r}{2}} - 1, N)$ divides $a^{\frac{r}{2}} -1$. 
If condition \eqref{eq: criteria2} is not satisfied (so that $N | a^{\frac{r}{2}} +1$), $p_1$ must also divide $a^{\frac{r}{2}} +1$, which is only possible if $p_1 \leq 2$ since $p_1$, $a$ and $r$ are integers. 
Furthermore, the relation $N | a^{\frac{r}{2}} +1$ implies $p_2=N$, a trivial factor. 
(The same argument can be applied to the condition  \eqref{eq: criteria1}.) 
Hence, for any interesting $N$, we consider the conditions \eqref{eq: criteria1} and \eqref{eq: criteria2} necessary and sufficient for finding the factors.  

The QOFA returns the trivial factors when these conditions are not satisfied. 
But the subject of this letter is the second cause of failure, odd values of $r$. 

\section{Factoring with odd orders}\label{sec: factoring with odd r}
Martin-Lopez \emph{et al.} \cite{eml-nphot-6-773} considered factoring $21$ with the coprime four giving order three. 
Despite the order being odd, the algorithm successfully returns the factors, $3=\gcd(4^{\frac{3}{2}} + 1, 21)$ and $7=\gcd(4^{\frac{3}{2}} - 1, 21)$, which are integer because the coprime is square. 

Square coprimes do not \emph{always} allow
this trick, however. 
Take factoring $21$ with coprime $16$. 
The order three leads to the trivial factors $\gcd(16^{\frac{3}{2}} + 1, 21) = 1$ and $\gcd(16^{\frac{3}{2}} - 1, 21) = 21$ because the condition  \eqref{eq: criteria1} is not met, $21 |  16^{\frac{3}{2}} -1$. 

So how often are the factors found from odd orders? 
To calculate this we must know the effect of square coprimes. 

We start with a definition. 
Let a square number, $b$, be written 
\begin{align}\label{eq: def b}
b = a^{2^m},
\end{align}
for positive integer $m$, in terms of a non-square \emph{root}, $a$. 

We define the order $s$ to be the smallest integer satisfying the order relation for coprime $b$, 
\begin{align}
b^s \mod N =1,\label{eq b order finding}
\end{align}
which, according to equation \eqref{eq: def b}, can be written $a^{2^m s} \mod N =1$. 
Clearly $a$ has its own order, $r$, 
\begin{align}
a^r \mod N =1.  \label{eq a order finding}
\end{align}
Since $r$ is optimal -- there is no smaller integer satisfying equation \eqref{eq a order finding} -- we have $r | 2^m s$. 
Without loss of generality we write $r= 2^n r_0$, where $r_0$ is odd, and hence, 
\begin{align}
 s =x 2^{n-m} r_0, 
\end{align}
where $x$ is the smallest positive integer such that $s$ is integer. The value of $x$ depends on $n$ and $m$. 

First, consider $n > m$, meaning that $r$ is even (since $m > 0$). 
In this case $x=1$ and $s= 2^{n-m} r_0$ (which is even) and so the order finding relation, 
\begin{align}
 b^{ 2^{n-m} r_0} \mod N =1,
\end{align}
is identical to equation \eqref{eq a order finding}; 
here, order finding with $b$ is the same process as order finding with $a$. 

Second, if $n=m$ (meaning $r$ is even) then, again, $x=1$, the order finding relation for $b$ is identical to that for $a$. But this time $s$ is odd, $s= r_0$. 

Finally, if $n <m$, then $x=2^{m-n}$ and $s=r_0$ (so $s$ is odd). 
Equation \eqref{eq b order finding} can be written in terms of $a$ and $r$, 
\begin{align}
 a^{r 2^{m-n}} Ê\mod N =1. 
\end{align}
$2^{m-n-1}$ is integer, so the condition \eqref{eq: criteria1} is not satisfied, $N | a^{r 2^{m-n-1}} -1$, and, thus, the factors are not found. 

Using coprime $b$, SFA gives the factors only in the first case -- when $s$ is even -- and only if $a$ also gives the factors. 
Considering odd orders improves this slightly. When $n=m$, the factors, $b^{\frac{r_0}{2}} \pm1$ are identical to those arising from the coprime $a$, $a^{\frac{2^n r_0}{2}}\pm1$, and so are found whenever they would have been found using $a$. 
This explains the calculation by reference \cite{eml-nphot-6-773}, where the coprime $b=4$ gave order $s=3$, equivalent to using $a=2$ as the coprime, giving order $r=6$ (here, $m=n=1$). \\

A second observation sometimes lets us retrieve the factors from a failed calculation.   
Factoring $N=21$ with the coprime $b=16$ fails because the condition \eqref{eq: criteria1} is not met. 
But this implies a new order finding relation, in terms of the coprime $a=4$, 
\begin{align}
 16^{\frac{3}{2}} \mod 21 \equiv 
 4^{3} \mod 21 = 1.
\end{align}
We have recovered the calculation of reference \cite{eml-nphot-6-773} which, of course, does satisfy the two conditions and leads to the factors, $\gcd(4^{\frac{3}{2}} \pm 1, 21)$. 
This works when $n <m$. Failing the condition \eqref{eq: criteria1}, $N | a^{r 2^{m-n-1}} -1$, implies an order finding relation for the coprime $\sqrt{b} = a^{2^{m-1}}$,
\begin{align}
a^{r 2^{m-n-1}}  \mod N =1. 
\end{align}
This process can be repeated;  
the factors are not found if $N | a^{r 2^{m-n-2}} -1$, in which case we have recovered the order finding relation for the coprime $b^{\frac{1}{4}} = a^{2^{m-2}}$. 
After $m-n$ repetitions we arrive at equation \eqref{eq a order finding}, and the problem is reduced to order finding with the \emph{root}, $a$.

\section{The effect on efficiency}
These two techniques -- considering odd orders and collapsing the coprime to its root -- 
let us find the factors from coprime $b$ and (odd) $s$ \emph{iff} (if and only if) the \emph{root} $a$ would have given them using the normal SFA procedure. 
They imply that the probability of factoring can be improved by considering only non-square coprimes. We now calculate this improvement. 

Let us consider SFA -- without excluding square coprimes -- for factoring $N = p_1 p_2$. 

Let the coprime $c$, picked uniformly at random ($1< c <N$), have order $t$, 
\begin{align}
c^t \mod N = 1. 
\end{align}
When $c$ is the (non-square) \emph{root} we will consider coprime $c=a$ (giving order $r$), otherwise we will use $c=b$ (with order $s$).

SFA finds the factors -- given $c$ and $t$ -- with probability
\begin{align}
&P (\text{factors}|c,t)\notag\\
& =  P(\text{factors}| a, r) P(c=a) + P(\text{factors}| b,s) P(c=b), \notag\\ 
& =  P(\text{factors}| a, r)\big(  P(c=a) + P(n>m)  P(c=b) \big), 
\end{align}
where $P$ means \emph{probability} ($P(c=a)$ is the probability that $c$ is non-square, for instance) and where we have used $P(\text{factors}| b,s) = P(\text{factors}| a, r) P(n>m)$ since the factors are found from $b$ only if $n >m$ and if they could have been found using the coprime $a$. 

Our techniques show that we can consider only non-square coprimes, which lead to the factors with probability $P(\text{factors}| a, r)$.
We compare this to the original, 
\begin{align}
\frac{P(\text{factors}|c,t)}{P(\text{factors}| a, r)} = P(c=a) + P(n>m)  P(c=b). 
\end{align}
But,  
\begin{align}
P(\text{$t$ even}) & \equiv P(n>0) P(c=a) + P(\text{$s$ even}) P(c=b)\notag\\
&\geq P(n>m)
\end{align}
because $P(\text{$s$ even}) = P(n>m)$ and $P(n>0) \geq P(n>m)$ since $m >0$, meaning that 
\begin{align}
\frac{P(\text{factors}|c,t)}{P(\text{factors}| a, r)} \leq  P(c=a) + P(\text{$t$ even})  P(c=b). 
\end{align}

The probability that $c$ is non-square is $P(c=a) = 1 - 1/\sqrt{N}$. This also defines $P(c=b)$ since $P(c=a) + P(c=b) = 1$. 
All that remains is to calculate the probability of $t$ being even. 
We assume that $p_i -1 = 2 q_i$, where $q_i$ are odd, corresponding to the hardest numbers to factor both quantumly, since it leads to lots of odd orders, and classically, using algorithms such as that proposed in reference \cite{Pollard1974}, which rely on $p_i$ being \emph{smooth}. 
Numbers of this form are thus likely candidates for SFA. 
In this case $P(\text{$t$ even}) = 3/4$, following the argument of reference \cite{nielsen+chuang2000}. For completeness we sketch the proof here. 

\begin{proof}
The Chinese remainder theorem (CRT) tell us that choosing $c$ uniformly at random from $1 < c < N$ is equivalent to randomly picking two integers, $c_1$ ($1< c_1 < p_1$) and $c_2$ ($1< c_2 < p_2$), where $c = c_i \mod p_i$. 
Let $t_i$ be the order satisfying 
\begin{align}
c_i ^{t_i} \mod p_i =1. 
\end{align}
According to the CRT this order finding relation is also satisfied by $t$, giving $t = \text{LCM}(t_1, t_2)$, where $\text{LCM}$ means \emph{least common multiple}. $t$ is odd only when both $t_1$ and $t_2$ are odd. How likely is this? 
Answering this is made easier by the fact that the multiplicative group $\!\!\!\! \mod p_1$ is cyclic. The elements of this group can be written in terms of a generator, $g$. Thus, $c_i \equiv g^k \mod p_i$ for some integer $k$ ($1 \leq k \leq p_i -1$). 
The order finding relation implies $g^{k t_i} \mod p_i=1$. 
But $g^{p_i-1} \mod p_i =1$, meaning that $p_i-1 | k t_i $. 
$p_i-1$ is even and so, if $k$ is odd, $t_i$ must be even. 
Alternatively, if $k$ is even, 
\begin{align}
g^{(p_i -1)\frac{k}{2}}\mod p_i = 1, 
\end{align}
so $t_i$ must be odd since it divides $(p_i -1)/2 = q_i$. 
The probability that $t_i$ is odd is therefore the probability that $k$ is even, which is $1/2$ since $c_i$ is picked at random. 
Hence, $P(\text{$t$ even}) = 1- P(t_1 \text{ odd}) P(t_2 \text{ odd}) = 3/4$. 
\end{proof}

In the best case, assuming $p_i -1= 2 q_i$, avoiding non-square coprimes improves the probability of success of SFA,  
\begin{align}
\frac{P(\text{factors}|c,t)}{P(\text{factors}| a, r)} \leq  1 - \frac{1}{4 \sqrt{N}}. 
\end{align}

\section{Discussion}

In this letter we correct a common misconception, showing that odd orders \emph{do} play a useful role in factoring, and should not be neglected outright. 
The existence of useful odd orders raises the question of improving the efficiency of SFA. We show that an improvement is possible, most simply by avoiding square coprimes. 
We are not the first to suggest this: 
Markov $\&$ Saeedi use numerical evidence to argue that SFA should use small prime coprimes like $a=2$, $3$ and $5$ \cite{Markov2012}; 
Leander showed how to avoid odd orders by picking coprimes that are non-square \emph{under modular arithmetic}, a property that can be efficiently checked using the Jacobi symbol \cite{Leander2002}. 
Here, we highlight another advantage of avoiding square coprimes: that a square and its root is never picked in different runs of the same calculation which, as we have shown, is a waste of resources; working through the coprimes in order, $a= 2$, $3$, $4$, $\ldots$, until the factors are found is certainly not efficient!

The improvement we propose is small, especially as $N$ becomes large. 
(Indeed, it is smaller than that proposed by Leander.) 
Nonetheless, the role of odd orders needed to be investigated to know that larger gains were not possible. Furthermore, for proof of principle experiments, odd orders may be desirable, especially in the near future, while young technologies 
restrict demonstrations to very small numbers. 
In this case, odd orders which lead to the factors are particularly attractive, since they are sure to avoid problematic power-of-two orders, $r=2^p$.

Quantum subroutines are -- and will probably remain for some time -- much harder to implement than classical ones. 
This is especially true of quantum circuits that use young, imperfect technologies, which introduce their own errors and hold-ups, and may need to be reconfigured each time the calculation changes. 
While this is the case small improvements in efficiency may give substantial savings in run-time.

As we have shown, even well studied algorithms like SFA are not fully understood.
With luck, a better knowledge of the algorithm will lead to better efficiency saving techniques, just as technological understanding has improved experimental demonstrations of the algorithm. Eventually, this 
will make experimentalists' lives easier, and bring about a convincing demonstration of SFA all the more quickly. \\


\noindent {\large \textbf{Acknowledgments}}

\noindent Thanks to Frederic Grosshans, Marc Kaplan, Anthony Laing, Marc-Andre Lajoie, Enrique Martin-Lopez and Benjamin Smith for valuable discussions. 
The author acknowledges support from Digiteo and the City of Paris project CiQWii.\\

\bibliography{Bib}

\begin{thebibliography}{13}%
\makeatletter
\providecommand \@ifxundefined [1]{%
 \@ifx{#1\undefined}
}%
\providecommand \@ifnum [1]{%
 \ifnum #1\expandafter \@firstoftwo
 \else \expandafter \@secondoftwo
 \fi
}%
\providecommand \@ifx [1]{%
 \ifx #1\expandafter \@firstoftwo
 \else \expandafter \@secondoftwo
 \fi
}%
\providecommand \natexlab [1]{#1}%
\providecommand \enquote  [1]{``#1''}%
\providecommand \bibnamefont  [1]{#1}%
\providecommand \bibfnamefont [1]{#1}%
\providecommand \citenamefont [1]{#1}%
\providecommand \href@noop [0]{\@secondoftwo}%
\providecommand \href [0]{\begingroup \@sanitize@url \@href}%
\providecommand \@href[1]{\@@startlink{#1}\@@href}%
\providecommand \@@href[1]{\endgroup#1\@@endlink}%
\providecommand \@sanitize@url [0]{\catcode `\\12\catcode `\$12\catcode
  `\&12\catcode `\#12\catcode `\^12\catcode `\_12\catcode `\%12\relax}%
\providecommand \@@startlink[1]{}%
\providecommand \@@endlink[0]{}%
\providecommand \url  [0]{\begingroup\@sanitize@url \@url }%
\providecommand \@url [1]{\endgroup\@href {#1}{\urlprefix }}%
\providecommand \urlprefix  [0]{URL }%
\providecommand \Eprint [0]{\href }%
\providecommand \doibase [0]{http://dx.doi.org/}%
\providecommand \selectlanguage [0]{\@gobble}%
\providecommand \bibinfo  [0]{\@secondoftwo}%
\providecommand \bibfield  [0]{\@secondoftwo}%
\providecommand \translation [1]{[#1]}%
\providecommand \BibitemOpen [0]{}%
\providecommand \bibitemStop [0]{}%
\providecommand \bibitemNoStop [0]{.\EOS\space}%
\providecommand \EOS [0]{\spacefactor3000\relax}%
\providecommand \BibitemShut  [1]{\csname bibitem#1\endcsname}%
\let\auto@bib@innerbib\@empty
\bibitem [{\citenamefont {Martin-Lopez}\ \emph {et~al.}(2012)\citenamefont
  {Martin-Lopez}, \citenamefont {Laing}, \citenamefont {Lawson}, \citenamefont
  {Alvarez}, \citenamefont {Zhou},\ and\ \citenamefont
  {O'Brien}}]{eml-nphot-6-773}%
  \BibitemOpen
  \bibfield  {author} {\bibinfo {author} {\bibfnamefont {E.}~\bibnamefont
  {Martin-Lopez}}, \bibinfo {author} {\bibfnamefont {A.}~\bibnamefont {Laing}},
  \bibinfo {author} {\bibfnamefont {T.}~\bibnamefont {Lawson}}, \bibinfo
  {author} {\bibfnamefont {R.}~\bibnamefont {Alvarez}}, \bibinfo {author}
  {\bibfnamefont {X.-Q.}\ \bibnamefont {Zhou}}, \ and\ \bibinfo {author}
  {\bibfnamefont {J.~L.}\ \bibnamefont {O'Brien}},\ }\href
  {http://dx.doi.org/10.1038/nphoton.2012.259} {\bibfield  {journal} {\bibinfo
  {journal} {Nat Photon}\ }\textbf {\bibinfo {volume} {6}},\ \bibinfo {pages}
  {773} (\bibinfo {year} {2012})}\BibitemShut {NoStop}%
\bibitem [{\citenamefont {Vandersypen}\ \emph {et~al.}(2001)\citenamefont
  {Vandersypen}, \citenamefont {Steffen}, \citenamefont {Breyta}, \citenamefont
  {Yannoni}, \citenamefont {Sherwood},\ and\ \citenamefont
  {Chuang}}]{Vandersypen2001}%
  \BibitemOpen
  \bibfield  {author} {\bibinfo {author} {\bibfnamefont {L.~M.~K.}\
  \bibnamefont {Vandersypen}}, \bibinfo {author} {\bibfnamefont
  {M.}~\bibnamefont {Steffen}}, \bibinfo {author} {\bibfnamefont
  {G.}~\bibnamefont {Breyta}}, \bibinfo {author} {\bibfnamefont {C.~S.}\
  \bibnamefont {Yannoni}}, \bibinfo {author} {\bibfnamefont {M.~H.}\
  \bibnamefont {Sherwood}}, \ and\ \bibinfo {author} {\bibfnamefont {I.~L.}\
  \bibnamefont {Chuang}},\ }\href {\doibase http://dx.doi.org/10.1038/414883a}
  {\bibfield  {journal} {\bibinfo  {journal} {Nature}\ }\textbf {\bibinfo
  {volume} {414}},\ \bibinfo {pages} {883} (\bibinfo {year}
  {2001})}\BibitemShut {NoStop}%
\bibitem [{\citenamefont {Lu}\ \emph {et~al.}(2007)\citenamefont {Lu},
  \citenamefont {Browne}, \citenamefont {Yang},\ and\ \citenamefont
  {Pan}}]{lu-prl-99-250504}%
  \BibitemOpen
  \bibfield  {author} {\bibinfo {author} {\bibfnamefont {C.-Y.}\ \bibnamefont
  {Lu}}, \bibinfo {author} {\bibfnamefont {D.~E.}\ \bibnamefont {Browne}},
  \bibinfo {author} {\bibfnamefont {T.}~\bibnamefont {Yang}}, \ and\ \bibinfo
  {author} {\bibfnamefont {J.-W.}\ \bibnamefont {Pan}},\ }\href {\doibase
  10.1103/PhysRevLett.99.250504} {\bibfield  {journal} {\bibinfo  {journal}
  {Phys. Rev. Lett.}\ }\textbf {\bibinfo {volume} {99}},\ \bibinfo {eid}
  {250504} (\bibinfo {year} {2007})}\BibitemShut {NoStop}%
\bibitem [{\citenamefont {Lanyon}\ \emph {et~al.}(2007)\citenamefont {Lanyon},
  \citenamefont {Weinhold}, \citenamefont {Langford}, \citenamefont {Barbieri},
  \citenamefont {James}, \citenamefont {Gilchrist},\ and\ \citenamefont
  {White}}]{Lanyon2007}%
  \BibitemOpen
  \bibfield  {author} {\bibinfo {author} {\bibfnamefont {B.~P.}\ \bibnamefont
  {Lanyon}}, \bibinfo {author} {\bibfnamefont {T.~J.}\ \bibnamefont
  {Weinhold}}, \bibinfo {author} {\bibfnamefont {N.~K.}\ \bibnamefont
  {Langford}}, \bibinfo {author} {\bibfnamefont {M.}~\bibnamefont {Barbieri}},
  \bibinfo {author} {\bibfnamefont {D.~F.~V.}\ \bibnamefont {James}}, \bibinfo
  {author} {\bibfnamefont {A.}~\bibnamefont {Gilchrist}}, \ and\ \bibinfo
  {author} {\bibfnamefont {A.~G.}\ \bibnamefont {White}},\ }\href {\doibase
  10.1103/PhysRevLett.99.250505} {\bibfield  {journal} {\bibinfo  {journal}
  {Phys. Rev. Lett.}\ }\textbf {\bibinfo {volume} {99}},\ \bibinfo {pages}
  {250505} (\bibinfo {year} {2007})}\BibitemShut {NoStop}%
\bibitem [{\citenamefont {Politi}\ \emph {et~al.}(2009)\citenamefont {Politi},
  \citenamefont {Matthews},\ and\ \citenamefont {O'Brien}}]{Mathews2009}%
  \BibitemOpen
  \bibfield  {author} {\bibinfo {author} {\bibfnamefont {A.}~\bibnamefont
  {Politi}}, \bibinfo {author} {\bibfnamefont {J.~C.~F.}\ \bibnamefont
  {Matthews}}, \ and\ \bibinfo {author} {\bibfnamefont {J.~L.}\ \bibnamefont
  {O'Brien}},\ }\href {\doibase 10.1126/science.1173731} {\bibfield  {journal}
  {\bibinfo  {journal} {Science}\ }\textbf {\bibinfo {volume} {325}},\ \bibinfo
  {pages} {1221} (\bibinfo {year} {2009})}\BibitemShut {NoStop}%
\bibitem [{\citenamefont {Lucero}\ \emph {et~al.}(2012)\citenamefont {Lucero},
  \citenamefont {Barends}, \citenamefont {Chen}, \citenamefont {Kelly},
  \citenamefont {Mariantoni}, \citenamefont {Megrant}, \citenamefont
  {O'Malley}, \citenamefont {Sank}, \citenamefont {Vainsencher}, \citenamefont
  {Wenner}, \citenamefont {White}, \citenamefont {Yin}, \citenamefont
  {Cleland},\ and\ \citenamefont {Martinis}}]{lu-nphys-10-719}%
  \BibitemOpen
  \bibfield  {author} {\bibinfo {author} {\bibfnamefont {E.}~\bibnamefont
  {Lucero}}, \bibinfo {author} {\bibfnamefont {R.}~\bibnamefont {Barends}},
  \bibinfo {author} {\bibfnamefont {Y.}~\bibnamefont {Chen}}, \bibinfo {author}
  {\bibfnamefont {J.}~\bibnamefont {Kelly}}, \bibinfo {author} {\bibfnamefont
  {M.}~\bibnamefont {Mariantoni}}, \bibinfo {author} {\bibfnamefont
  {A.}~\bibnamefont {Megrant}}, \bibinfo {author} {\bibfnamefont
  {P.}~\bibnamefont {O'Malley}}, \bibinfo {author} {\bibfnamefont
  {D.}~\bibnamefont {Sank}}, \bibinfo {author} {\bibfnamefont {A.}~\bibnamefont
  {Vainsencher}}, \bibinfo {author} {\bibfnamefont {J.}~\bibnamefont {Wenner}},
  \bibinfo {author} {\bibfnamefont {T.}~\bibnamefont {White}}, \bibinfo
  {author} {\bibfnamefont {Y.}~\bibnamefont {Yin}}, \bibinfo {author}
  {\bibfnamefont {A.~N.}\ \bibnamefont {Cleland}}, \ and\ \bibinfo {author}
  {\bibfnamefont {J.~M.}\ \bibnamefont {Martinis}},\ }\href
  {http://dx.doi.org/10.1038/nphys2385} {\bibfield  {journal} {\bibinfo
  {journal} {Nat Phys}\ }\textbf {\bibinfo {volume} {8}},\ \bibinfo {pages}
  {719} (\bibinfo {year} {2012})}\BibitemShut {NoStop}%
\bibitem [{\citenamefont {Nielsen}\ and\ \citenamefont
  {Chuang}(2000)}]{nielsen+chuang2000}%
  \BibitemOpen
  \bibfield  {author} {\bibinfo {author} {\bibfnamefont {M.~A.}\ \bibnamefont
  {Nielsen}}\ and\ \bibinfo {author} {\bibfnamefont {I.~L.}\ \bibnamefont
  {Chuang}},\ }\href@noop {} {\emph {\bibinfo {title} {Quantum Computation and
  Quantum Information}}}\ (\bibinfo  {publisher} {Cambridge U.P.},\ \bibinfo
  {year} {2000})\BibitemShut {NoStop}%
\bibitem [{\citenamefont {{Leander}}(2002)}]{Leander2002}%
  \BibitemOpen
  \bibfield  {author} {\bibinfo {author} {\bibfnamefont {G.}~\bibnamefont
  {{Leander}}},\ }\href@noop {} {\bibfield  {journal} {\bibinfo  {journal}
  {arXiv: 0208183 [quant-ph]}\ } (\bibinfo {year} {2002})}\BibitemShut
  {NoStop}%
\bibitem [{\citenamefont {Markov}\ and\ \citenamefont
  {Saeedi}(2012)}]{Markov2012}%
  \BibitemOpen
  \bibfield  {author} {\bibinfo {author} {\bibfnamefont {I.~L.}\ \bibnamefont
  {Markov}}\ and\ \bibinfo {author} {\bibfnamefont {M.}~\bibnamefont
  {Saeedi}},\ }\href {http://dl.acm.org/citation.cfm?id=2230996.2230997}
  {\bibfield  {journal} {\bibinfo  {journal} {Quantum Info. Comput.}\ }\textbf
  {\bibinfo {volume} {12}},\ \bibinfo {pages} {361} (\bibinfo {year}
  {2012})}\BibitemShut {NoStop}%
\bibitem [{\citenamefont {Shor}(1997)}]{Shor1994}%
  \BibitemOpen
  \bibfield  {author} {\bibinfo {author} {\bibfnamefont {P.~W.}\ \bibnamefont
  {Shor}},\ }\href@noop {} {\bibfield  {journal} {\bibinfo  {journal} {SIAM J.
  Sci. Statist. Comput.}\ }\textbf {\bibinfo {volume} {26}},\ \bibinfo {pages}
  {1484} (\bibinfo {year} {1997})}\BibitemShut {NoStop}%
\bibitem [{\citenamefont {Knill}(1995)}]{Knill1995}%
  \BibitemOpen
  \bibfield  {author} {\bibinfo {author} {\bibfnamefont {E.}~\bibnamefont
  {Knill}},\ }\href@noop {} {\enquote {\bibinfo {title} {On shor's quantum
  factor finding algorithm: Increasing the probability of success and tradeoffs
  involving the fourier transform modulus},}\ }\bibinfo {howpublished} {Tech.
  Report LAUR-95-3350, Los Alamos Natl. Lab} (\bibinfo {year}
  {1995})\BibitemShut {NoStop}%
\bibitem [{\citenamefont {{Smolin}}\ \emph {et~al.}(2013)\citenamefont
  {{Smolin}}, \citenamefont {{Smith}},\ and\ \citenamefont
  {{Vargo}}}]{Smolin2013}%
  \BibitemOpen
  \bibfield  {author} {\bibinfo {author} {\bibfnamefont {J.~A.}\ \bibnamefont
  {{Smolin}}}, \bibinfo {author} {\bibfnamefont {G.}~\bibnamefont {{Smith}}}, \
  and\ \bibinfo {author} {\bibfnamefont {A.}~\bibnamefont {{Vargo}}},\
  }\href@noop {} {\bibfield  {journal} {\bibinfo  {journal} {Nature}\ }\textbf
  {\bibinfo {volume} {499}},\ \bibinfo {pages} {163} (\bibinfo {year}
  {2013})}\BibitemShut {NoStop}%
\bibitem [{\citenamefont {{Pollard}}(1974)}]{Pollard1974}%
  \BibitemOpen
  \bibfield  {author} {\bibinfo {author} {\bibfnamefont {J.~M.}\ \bibnamefont
  {{Pollard}}},\ }\href@noop {} {\bibfield  {journal} {\bibinfo  {journal}
  {Mathematical Proceedings of the Cambridge Philosophical Society}\ }\textbf
  {\bibinfo {volume} {76}},\ \bibinfo {pages} {3049} (\bibinfo {year}
  {1974})}\BibitemShut {NoStop}%
\end{thebibliography}%

\end{document}